\documentclass[aps,pra,showkeys,twocolumn,superscriptaddress]{revtex4-2}
\usepackage{array}
\usepackage{booktabs}
\usepackage{tabu}
\usepackage{dcolumn}
\usepackage{floatrow}
\usepackage{amsmath}
\usepackage{amsfonts}
\usepackage{float}
\usepackage{amssymb}
\usepackage{graphicx,color}
\usepackage[colorlinks={true}]{hyperref}
\hypersetup{colorlinks=true,linkcolor=red,citecolor=blue,urlcolor=blue}
\usepackage{graphicx}
\usepackage{dcolumn}
\usepackage{bm}
\usepackage{pstricks}
\usepackage{braket}
\usepackage{orcidlink}
\usepackage{lipsum}
\def\be{\begin{equation}}
  \def\ee{\end{equation}}
\def\bea{\begin{eqnarray}}
\def\eea{\end{eqnarray}}
\def\f{\frac}
\def\n{\nonumber}
\def\l{\label}
\def\p{\phi}
\def\o{\over}
\def\R{\rho}
\def\pa{\partial}
\def\om{\omega}
\def\na{\nabla}
\def\P{\Phi}
\begin{document}

\title{Work Extraction from Classically Correlated States in Noisy Quantum Channels with Memory} 

\author{Maryam Hadipour \orcidlink{0000-0002-6573-9960}}
\affiliation{Faculty of Physics, Urmia University of Technology, Urmia, Iran}

\author{Soroush Haseli \orcidlink{0000-0003-1031-4815}}\email{soroush.haseli@uut.ac.ir}
\affiliation{Faculty of Physics, Urmia University of Technology, Urmia, Iran}
\affiliation{School of Quantum Physics and Matter, Institute for Research in Fundamental Sciences (IPM), P.O. 19395-5531, Tehran, Iran}


\date{\today}
\def\be{\begin{equation}}
  \def\ee{\end{equation}}
\def\bea{\begin{eqnarray}}
\def\eea{\end{eqnarray}}
\def\f{\frac}
\def\n{\nonumber}
\def\l{\label}
\def\p{\phi}
\def\o{\over}
\def\R{\rho}
\def\pa{\partial}
\def\om{\omega}
\def\na{\nabla}
\def\P{$\Phi$}

\begin{abstract}
This study investigates the potential of local non-unital noise and quantum channel memory to enhance work extraction from classically correlated quantum states. Utilizing the framework of daemonic ergotropy, which incorporates measurement-based feedback via an ancillary system, we show that amplitude damping channels can induce quantum correlations that enable additional extractable work. Through analytical derivations and numerical simulations, we quantify the daemonic gain and demonstrate that channel memory significantly amplifies this advantage by preserving system-ancilla correlations. Our results reveal that  non-unital noise can serve not as a limitation but as a valuable thermodynamic resource in quantum protocols.

\end{abstract}

\maketitle

\section{Introduction}\label{sec1}
 Quantum thermodynamics stands at the forefront of understanding how fundamental thermodynamic principles extend into the quantum realm. It lays the groundwork for analyzing and exploiting the relationships between energy, entropy, coherence, and information in quantum systems \cite{1,2,3}. This field provides a theoretical framework for exploring the interplay between energy, information, and coherence at the quantum level. Understanding these interactions is essential for the effective implementation of processes such as work extraction in quantum systems. Therefore, quantum thermodynamics plays a crucial role in the design and optimization of next-generation quantum devices.

In recent years, quantum thermodynamics has played a pivotal role in the design of nano- and microscale engines with remarkable efficiency. The unique properties of quantum systems have enabled the development of protocols that manipulate and control work and heat in ways that surpass the limits of classical thermodynamics. These advancements have not only laid the foundation for a coherent framework in quantum caloritronics \cite{4,5,6}, but have also led to the experimental realization of Maxwell's demon using simple quantum systems as working media \cite{7,8}. These achievements establish a clear connection between theoretical concepts and practical applications in the field of quantum technologies.

Ergotropy is defined as the maximum amount of work that can be extracted from a quantum system via unitary operations, without altering the system's entropy \cite{9}. It quantifies the useful energy stored in a non-passive quantum state—i.e., a state from which work can be extracted through reversible dynamics \cite{9,10,11,12,13,14,15,16,17,18,19}. In contrast, passive states are thermodynamically inert under unitary evolution, despite possibly having non-zero internal energy.

In recent years, significant attention has been devoted to the study of ergotropy within quantum thermodynamics. Numerous works have explored its relationship to quantum coherence, correlations, and the structure of open-system dynamics, particularly in the context of finite systems, quantum engines, and measurement-assisted protocols \cite{z1,z2,z3,z4}. These investigations have helped clarify how ergotropy serves as a reliable metric for evaluating the energetic potential of quantum states and for assessing the operational role of quantum resources.

In the quantum regime, ergotropy becomes especially relevant due to its sensitivity to state ordering and its distinction from classical notions of energy and entropy. It can remain zero even for energetic states if the populations are thermodynamically ordered (passive), and it can increase due to the presence of quantum coherence or ancilla-assisted feedback \cite{Francica2017}.

Quantum coherence has been identified as a key enabler for extracting work from a single thermal reservoir \cite{20} and for enhancing the operational efficiency of quantum heat engines\cite{21}. Furthermore, it has been demonstrated that weakly driven quantum heat engines can outperform their classical (stochastic) counterparts in terms of power output \cite{22}.

Although various pieces of evidence have been presented, there is still no clear consensus on which specific features of quantum systems influence their thermodynamic performance. In particular, differing perspectives remain regarding the role that quantum discord and coherences play in the process of work extraction from quantum systems, and no unified theoretical agreement has yet been reached in this regard \cite{23,24,25,26,27,28}. While some works affirm that such correlations can act as operational resources, others suggest that entanglement is not strictly necessary for optimal performance \cite{25}. Given the ambiguity surrounding these claims, it is essential to develop frameworks that can accurately and delicately identify and harness quantum features within thermodynamic protocols.

Among the emerging approaches, daemonic ergotropy stands out as a particularly promising concept, first introduced in \cite{Francica2017}. This framework employs measurement-based feedback to enhance the efficiency of work extraction from quantum systems. By extending the concept of ergotropy through the inclusion of ancillary systems and the implementation of conditional operations based on measurement outcomes, it enables access to energy that would otherwise remain unattainable. Such methodologies not only reveal the deep interplay between quantum information and thermodynamic laws, but also highlight the crucial role of measurement and control in optimizing energy utilization at the quantum scale.

One of the emerging and intellectually stimulating directions within the framework of quantum information theory is the deliberate and constructive utilization of environmental noise. Contrary to the traditional perspective that treats noise as a disruptive and undesirable element, recent studies have demonstrated that noise can give rise to quantum correlations even from initially classically correlated states (particularly in the form of local non-unital quantum channels) \cite{Streltsov2011,abad}. This unexpected behavior challenges conventional intuitions regarding the role of noise in quantum systems and opens up a new paradigm in which noise can be engineered as a functional and purposeful resource in quantum information processing, rather than merely being suppressed or eliminated.

In this work, we explore how non-unital local noise and the memory effects of quantum channel can be harnessed for enhanced work extraction from classically correlated quantum states. By considering the amplitude damping channel as an archetype of dissipative, non-unital processes, we demonstrate that such noise can induce quantum correlations that contribute positively to thermodynamic performance. Furthermore, we incorporate memory effects by modeling correlated amplitude damping channels and show that channel memory serves to preserve or amplify quantum correlations across successive uses, thereby boosting the effectiveness of daemonic feedback protocols.
Our findings clearly indicate that although ergotropy alone exhibits limited performance under noisy conditions, the implementation of projective measurements on an ancillary system (ancilla) can lead to significantly enhanced work extraction. This gain becomes even more pronounced in the presence of channel memory, where it plays a more decisive role.

By employing both analytical methods and numerical simulations, we have evaluated the magnitude of daemonic ergotropy and its advantage over conventional ergotropy across a wide range of noise types and memory strengths. The results demonstrate that non-unital noise, not only ceases to be a hindrance but can also serve as a valuable thermodynamic resource in feedback-assisted quantum protocols, substantially boosting energy efficiency.
The work is organized as follows: In Sec. \ref{sec2}, we review the fundamental concept of ergotropy and its extension to daemonic ergotropy, highlighting the role of ancilla-assisted feedback in enhancing work extraction. In Sec. \ref{sec3}, we examine how local non-unital noise, particularly amplitude damping channels, can generate quantum correlations from classically correlated states and thereby increase the extractable work. In Sec. \ref{sec4}, we analyze the impact of quantum channel memory on work extraction, demonstrating that memory effects preserve system-ancilla correlations and enhance the effectiveness of daemonic protocols. Finally, in Sec. \ref{sec5}, we summarize our key results and discuss their implications for quantum thermodynamic resource theory and energy-efficient quantum technologies.
\section{Ancilla-Assisted Work Extraction: From Ergotropy to Daemonic Gain}\label{sec2}
In the field of quantum thermodynamics, investigating the impact of quantum correlations on work extraction from a system has opened new horizons for enhancing energy efficiency at the microscopic scale. This section introduces the concepts of ergotropy, daemonic ergotropy, and the resulting daemonic gain, accompanied by comprehensive theoretical explanations and relevant formulas.
\subsection{Ergotropy}
Under a fixed Hamiltonian $ \hat{H}^S$, the maximum amount of work that can be extracted from a quantum system in the state $ \hat{\rho}^S$ through cyclic unitary operations is known as \textit{ergotropy} \cite{20}. It quantifies the portion of the system's energy that is operationally accessible and can be harnessed for useful work. Let the quantum system be described by a time-independent Hamiltonian

\begin{equation}
\hat{H}^S = \sum_k \epsilon_k \vert \epsilon_k \rangle \langle \epsilon_k \vert,
\end{equation}

where $ \epsilon_k $ are the eigenvalues (energy levels) of the system, arranged in non-decreasing order $\epsilon_1 \leq \epsilon_2 \leq \cdots \leq \epsilon_n$. The state of the system is given by the density operator $ \hat{\rho}^S $, which can be written as

\begin{equation}
\hat{\rho}^S = \sum_k r_k \vert r_k \rangle \langle r_k \vert,
\end{equation}
where \( r_k \) are the eigenvalues (populations), ordered non-increasingly $r_1 \geq r_2 \geq \cdots \geq r_n$.

The ergotropy, denoted by $\mathcal{W}$, is defined as the difference between the actual average energy of the state and the minimal average energy achievable by any unitary transformation $U$ \cite{20}

\begin{equation}
\mathcal{W}= \mathrm{Tr}[\hat{\rho}^S \hat{H}^S] - \min_{U \in \mathcal{U}} \mathrm{Tr}[U \hat{\rho}^S U^\dagger \hat{H}^S],
\end{equation}
where $\mathcal{U}$ is the set of all unitary operators acting on the system's Hilbert space. An important consequence of the concept of ergotropy is that not all quantum states possess the capability to yield work. When the density matrix $\hat{\rho}_S$ is diagonal in the same basis as the Hamiltonian $\hat{H}_S$ and the populations $r_k$ increase monotonically with the energy levels $\epsilon_k$, no unitary operation exists that can further reduce the energy of the system . Such a state is referred to as a passive state $\hat{\sigma}_S=\sum_k r_k \vert \epsilon_k \rangle \langle \epsilon_k \vert$ and is characterized by the following properties
\begin{equation}
[\hat{\sigma}^S, \hat{H}^S] = 0  \Rightarrow \mathcal{W} = 0.
\end{equation}
Conversely, non-passive states can be partially transformed into passive states by a suitable unitary operation, allowing extraction of work in the process. In summary, the maximum amount of work that can be extracted can be expressed as
\begin{equation}
\mathcal{W}=\operatorname{Tr}\left(H_S \rho_S\right)-\operatorname{Tr}\left(H_S \sigma_S\right)
\end{equation}
From above, a more general expression for ergotropy can be obtained as \cite{20}
\begin{equation}\label{ergotropy}
\mathcal{W} = \sum_{j,k} r_k \epsilon_j \left( \vert \langle \epsilon_j \vert r_k \rangle \vert^2 - \delta_{jk} \right),
\end{equation}
\subsection{Daemonic Ergotropy}

Daemonic ergotropy extends the concept of ergotropy by incorporating feedback control assisted by an ancillary system, analogous to the operation of Maxwell's demon \cite{Francica2017}. In this framework, an ancillary system $ A$, which is correlated with the primary system $S$, is measured, and the outcomes of this measurement are used to perform conditional operations on $S$. This approach enables an enhanced extraction of work beyond what is achievable without such feedback.

The protocol begins with a joint bipartite state $\hat{\rho}^{SA}$ defined over the system $S$ and the ancilla $A$, with the assumption that there is no interaction Hamiltonian coupling the two subsystems during the process. A projective measurement described by a complete set of projectors $\{ \Pi_a^A \}$ is performed on the ancilla. Upon obtaining the measurement outcome $a$, the state of the system $S$ collapses conditionally to
\begin{equation}
\hat{\rho}_{S|_a} = \frac{\mathrm{Tr}_A \left[ \left( I_S \otimes \Pi_a^A \right) \hat{\rho}^{SA} \left( I_S \otimes \Pi_a^A \right) \right]}{p_a},
\end{equation}
where $p_a$ is the probability of obtaining outcome $a$, given by
\begin{equation}
p_a = \mathrm{Tr} \left[ \left( I_S \otimes \Pi_a^A \right) \hat{\rho}^{SA} \right].
\end{equation}
Following this, a conditional unitary operation $ U_a $ is applied to the system $S$ with the goal of maximizing the extractable work from the post-measurement conditional state $\hat{\rho}^S|_a$. The daemonic ergotropy is then defined as the average maximum work that can be extracted using this measurement and feedback protocol
\begin{equation}
\mathcal{W}_{\{ \Pi_a^A \}} = \mathrm{Tr}[\hat{\rho}^S \hat{H}^S] - \sum_a p_a \min_{U_a} \mathrm{Tr} \left[ U_a \hat{\rho}_{S|_a} U_a^\dagger \hat{H}^S \right].
\end{equation}
Equivalently, this quantity can be expressed as the weighted average of the ergotropy values of the conditional states \cite{Francica2017}
\begin{equation}
\mathcal{W}_{ \{ \Pi_a^A \}} = \sum_a p_a \mathcal{W}_a,
\end{equation}
where $ \mathcal{W}_a $ denotes the ergotropy of the conditional state $\hat{\rho}_{S|_a}$. This framework highlights the crucial role of measurement and feedback control in enhancing work extraction from quantum systems, underscoring the importance of information and quantum correlations in improving the efficiency of microscopic energy devices.

Beyond the rather extreme scenario discussed previously, there may exist other cases in which no increase in work extraction is observed despite the presence of correlations in the state $\hat{\rho}^{SA}$. The main objective here is to identify and characterize such instances. For this purpose a quantity termed the daemonic gain, has introduced as  follows \cite{Francica2017}

\begin{equation}
\delta \mathcal{W} = \max_{\{ \hat{\Pi}_a^A \}} \left(  \mathcal{W}_{\{ \hat{\Pi}_a^A \}} -  \mathcal{W} \right).
\end{equation}
This quantity represents the maximal enhancement achievable through optimization over the set of measurements \( \{ \hat{\Pi}_a^A \} \) performed on the ancilla. By definition and due to the optimization involved, it is evident that \(\delta W\) is always non-negative.
\section{Role of Local Non-Unital Channels in Enhancing Daemonic Ergotropy}\label{sec3}
Quantum correlations, even in the absence of entanglement, are recognized as fundamental resources in quantum information processing. Among these, quantum discord represents a form of non-classical correlation that can exist even in separable states. A study by Streltsov et al. \cite{Streltsov2011} has shown that local noise can, under certain conditions, give rise to quantum correlations in multi-qubit systems. Specifically, it has been demonstrated that non-unital quantum channels, when applied to classically correlated states, have the capacity to generate such correlations\cite{Streltsov2011,abad}.

This finding establishes a conceptual bridge between the dynamics of open quantum systems and quantum thermodynamics. In this work, we investigate how these noise-induced quantum correlations can be practically harnessed to optimize the process of work extraction. This analysis is conducted within the framework of Daemonic ergotropy, as introduced in \cite{Francica2017}. Following the framework of \cite{Streltsov2011}, we consider a bipartite system-ancilla state $\rho_{SA}$  that is fully classically correlated
\begin{equation}
\rho_{SA} = \sum_{i,j} p_{i,j} \vert i \rangle_S \langle i \vert  \otimes \vert j \rangle_A \langle j \vert 
\end{equation}
where $\vert i \rangle_S$ and $\vert j \rangle_j$ are orthonormal bases of parties $S$ and $A$, respectively, and $p_{ij} \geq 0$ with $\sum_{i,j} p_{ij} = 1$. These states have no quantum discord and are diagonal in a product basis. Now, we consider a bipartite quantum system composed of a system qubit $S$ and an ancilla qubit $A$. The initial state is fully classically correlated and given by
\begin{equation}\label{cc state}
\rho_{SA} = \frac{1}{2}\vert e \rangle_S \langle e \vert \otimes \vert e \rangle_A \langle e \vert + \frac{1}{2}\vert g \rangle_S \langle g \vert \otimes \vert g \rangle_A \langle g \vert.
\end{equation}
where $\vert e \rangle$ and $\vert g \rangle$ are excited and ground state, respectively. and  This state is diagonal in a product basis and contains no quantum discord, it belongs to the set of classical-correlated (CC) states. A bipartite quantum state is said to be quantum correlated if it cannot be written in the classically correlated form.

To investigate the emergence and impact of quantum correlations, we now focus on the effect of local quantum operations, specifically quantum channels. A quantum channel is a completely positive, trace-preserving (CPTP) map that describes the most general evolution of a quantum system, including noise and open system dynamics. Formally, any such channel $\lambda$ acting on a state $\rho$ can be expressed using its Kraus representation

\begin{equation}
\Lambda(\rho) = \sum_i K_i \rho K_i^\dagger,
\end{equation}
where the set of Kraus operators $\{K_i\}$ satisfies the completeness condition $\sum_i K_i^\dagger K_i = I$. This framework enables a rigorous analysis of how local operations influence quantum states and the correlations between their subsystems. Within the extensive variety of quantum channels, a fundamental classification arises based on how they act on the maximally mixed state. Specifically, one can categorize channels as follows

\begin{itemize}
  \item \textbf{Unital Channels:} These are maps that leave the maximally mixed state invariant. That is, for a system of dimension $d$, a unital channel $\Lambda$ satisfies
  \[
  \Lambda\left( \frac{I}{d} \right) = \frac{I}{d}.
  \]

  \item \textbf{Non-Unital Channels:} In contrast, these channels alter the maximally mixed state, so the relation
  \[
  \Lambda\left( \frac{I}{d} \right) \neq \frac{I}{d}
  \]
  holds. Such transformations are characteristic of dissipative or non-equilibrium processes.
\end{itemize}
This distinction plays a key role in understanding how quantum information evolves, especially in open system dynamics. A pivotal result, presented as theorem(1) in Ref.\cite{Streltsov2011}, establishes that the generation of quantum correlations through local operations is restricted to a specific class of quantum channels. In particular, it is only those channels that are both non-unital and non-semi-classical that possess the capacity to induce quantum correlations from states that are initially classically correlated. This foundational insight directly informs our selection of quantum channels for the subsequent analysis.

To exemplify the behavior of non-unital quantum channels, we focus on the amplitude damping channel, a prototypical model for irreversible processes involving energy loss, such as spontaneous emission. This channel captures the essential physics of relaxation to a lower energy state and is characterized by the following set of Kraus operators:

\begin{equation}\label{kramp}
K_0 = 
\begin{pmatrix}
1 & 0 \\
0 & \sqrt{1 - \gamma}
\end{pmatrix},
\qquad
K_1 = 
\begin{pmatrix}
0 & \sqrt{\gamma} \\
0 & 0
\end{pmatrix},
\end{equation}
where $\gamma \in [0,1]$ denotes the damping parameter, representing the probability of energy decay from the excited to the ground state. A noteworthy feature of the amplitude damping channel is its action on the maximally mixed state
\begin{equation}
\Lambda_{A D}\left(\frac{\mathbb{I}}{2}\right)=\frac{1}{2}\left(\begin{array}{cc}
1+\gamma & 0 \\
0 & 1-\gamma
\end{array}\right) \neq \frac{\mathbb{I}}{2}.
\end{equation}
From above, it can be seen that rather than preserving this state, the channel transforms it into a non-uniform, or biased, state, highlighting its non-unital nature and its tendency to favor lower energy populations. When the amplitude damping channel $\Lambda_{\mathrm{AD}}$ is applied locally to the system component of a bipartite state that is initially classically correlated with an ancilla, it can give rise to genuine quantum correlations in the evolved joint state. Importantly, these emergent correlations are not merely of academic interest. As we demonstrate in subsequent sections, they yield tangible benefits in practical settings, most notably by enhancing the extractable work from the system. This enhancement is realized via a feedback-assisted thermodynamic protocol, and is quantitatively captured by the concept of \textit{daemonic ergotropy}. 
To analyze the dynamical effects of local noise, we  apply the amplitude damping channel $\Lambda_{\mathrm{AD}}$ exclusively to the system qubit $S$, while keeping the ancilla qubit untouched. This localized operation results in the transformation of the initial bipartite state into a new joint state, given by

\begin{equation}
\rho_{SA}' = (\Lambda_{\mathrm{AD}} \otimes \mathbb{I})[\rho_{SA}],
\end{equation}
where $\rho_{SA}$ denotes the original system-ancilla state and $\mathbb{I}$ represents the identity map acting on the ancilla. By consider initial bipartite classical correlated state in Eq.\ref{cc state} as initial state the final state after apply the local amplitude damping channel is obtained as 
\begin{eqnarray}
\rho_{SA}' &=& \frac{1}{2} \vert e \rangle_S \langle e \vert \otimes \vert e \rangle_A \langle e \vert + \frac{\gamma}{2} \vert e \rangle_S \langle e \vert \otimes \vert g \rangle_A \langle g \vert \nonumber \\
&+& \frac{1-\gamma}{2} \vert g \rangle_S \langle g \vert \otimes \vert g \rangle_A \langle g \vert. 
\end{eqnarray}
The resulting output state is no longer diagonal in any product basis, signaling the presence of genuine quantum correlations as soon as $\gamma > 0$. These correlations arise from the action of the Kraus operator $K_1$, which induces population transfer between energy levels in a way that disrupts the classical structure of the initial state.

This emergent non-classicality carries significant operational implications. By establishing correlations between the ancilla and the system, the measurement outcomes on the ancilla become a valuable source of information. Specifically, such measurement results can be used to condition subsequent operations performed on the system, thereby enabling a conditional enhancement of thermodynamic efficiency. Within the framework of our study, this phenomenon is reflected in the improved extractable work, which is quantitatively captured by the concept of daemonic ergotropy. 

Prior to the application of the amplitude damping channel, the system resides in a maximally mixed state, which implies that no work can be extracted. In this state, the system exhibits neither asymmetry nor population imbalance, both of which are necessary resources for work extraction. Moreover, although the bipartite system-ancilla state contains classical correlations, these do not enhance the amount of work that can be obtained from the system alone. After applying the amplitude damping channel, the reduced system state becomes
\begin{equation}\label{reds}
\rho_S'=\frac{1+\gamma}{2}  \vert e \rangle_S\langle e \vert + \frac{1-\gamma}{2} \vert g \rangle_S \langle g \vert .
\end{equation}
Consider the system Hamiltonian defined by $H_S=\omega \sigma_+\sigma_-$, where $\sigma_+=\vert e \rangle \langle g \vert$, $\sigma_-=\vert g \rangle \langle e \vert$ and here, $\omega$ denotes the energy difference between the excited state $\vert e \rangle$ and the ground state $\vert g \rangle$. the expected energy of the system is computed as $\mathcal{E}_S = \mathrm{Tr}[\rho_S H_S] = \frac{1+\gamma}{2} \, \omega$. Due to the population of the excited state, $\frac{1+\gamma}{2}$, is greater than or equal to that of the ground state, $\frac{1-\gamma}{2}$, the state $\rho_S$ is non-passive. So, according to the definition of ergotropy given in Eq.\ref{ergotropy}, the ergotropy of the state is $\mathcal{W}=\gamma \, \omega$. After the amplitude damping process, the system exhibits a nonzero ergotropy that scales linearly with the parameter $\gamma$. This increase corresponds to the enhanced population of the excited state induced by the damping dynamics within the specified basis. 

However, daemonic ergotropy allows for enhanced work extraction through measurement-based feedback. Specifically, we consider a set of projective measurements $\{ \Pi_a^A\}$ on the ancillary qubit $A$, which will collapse the system into conditional states based on the outcome of the measurement $a$. We are now considering a measurement in the general basis defined by two orthonormal states
\begin{eqnarray}
\vert \psi_0 \rangle &=& \cos\left(\frac{\theta}{2}\right) \vert e \rangle + e^{i\phi} \sin\left(\frac{\theta}{2}\right) \vert g \rangle, \nonumber \\
\vert \psi_1 \rangle &=& -\sin \left(\frac{\theta}{2}\right) \vert e \rangle + e^{-i\phi} \cos\left(\frac{\theta}{2}\right) \vert g \rangle,
\end{eqnarray}
The corresponding projectors are
\begin{equation}\label{pms}
\Pi_0^A= \vert \psi_0 \rangle \langle \psi_0 \vert, \quad \Pi_1^A= \vert \psi_1 \rangle \langle \psi_1 \vert.
\end{equation}
Upon performing a projective measurement on the ancilla, the system undergoes a collapse to conditional states, determined by the outcome of the measurement. For the general projective measurement, the conditional states $\rho_S|0$ and $\rho_S|1$ can be obtained by tracing out the ancilla qubit from the total system state after the measurement process. The conditional state of the system following the measurement outcome $0$ is given by
\begin{equation}
\rho_{S|0} =(\cos^2 \theta/2 + \gamma \sin^2 \theta/2)\vert e \rangle \langle e \vert + (1-\gamma)\sin^2 \theta/2 \vert g \rangle \langle g \vert.
\end{equation}
Similarly, the conditional state of the system after the measurement outcome $1$ is
\begin{equation}
\rho_{S|1} =(\sin^2 \theta/2 + \gamma \cos^2 \theta/2)\vert e \rangle \langle e \vert + (1-\gamma) \cos^2 \theta/2 \vert g \rangle \langle g \vert,
\end{equation}

Once the conditional states $\rho_{S|0}$ and $\rho_{S|1}$ have been computed, we proceed to calculate the ergotropy for each of these states. The ergotropy \( \mathcal{W}(\rho_{S|a}) \) of any state \( \rho_{S|a} \) is given by
\begin{equation}
\mathcal{W}(\rho_{S|a})=\chi \cdot \Theta \left( \chi \right),
\end{equation}
where  is the Heaviside function $\chi= \gamma+(-1)^a(1-\gamma) \cos \theta $.
Finally, the total daemonic ergotropy is the weighted sum of the ergotropies from both measurement outcomes, with probabilities \( p_0=1/2 \) and \( p_1=1/2 \), given by

\begin{equation}
\mathcal{W}_{\{\Pi_A\}} = p_0 \mathcal{W}(\rho_S|0) + p_1 \mathcal{W}(\rho_S|1).
\end{equation}

\begin{figure}[t]
	\centering
	\includegraphics[width =1 \linewidth]{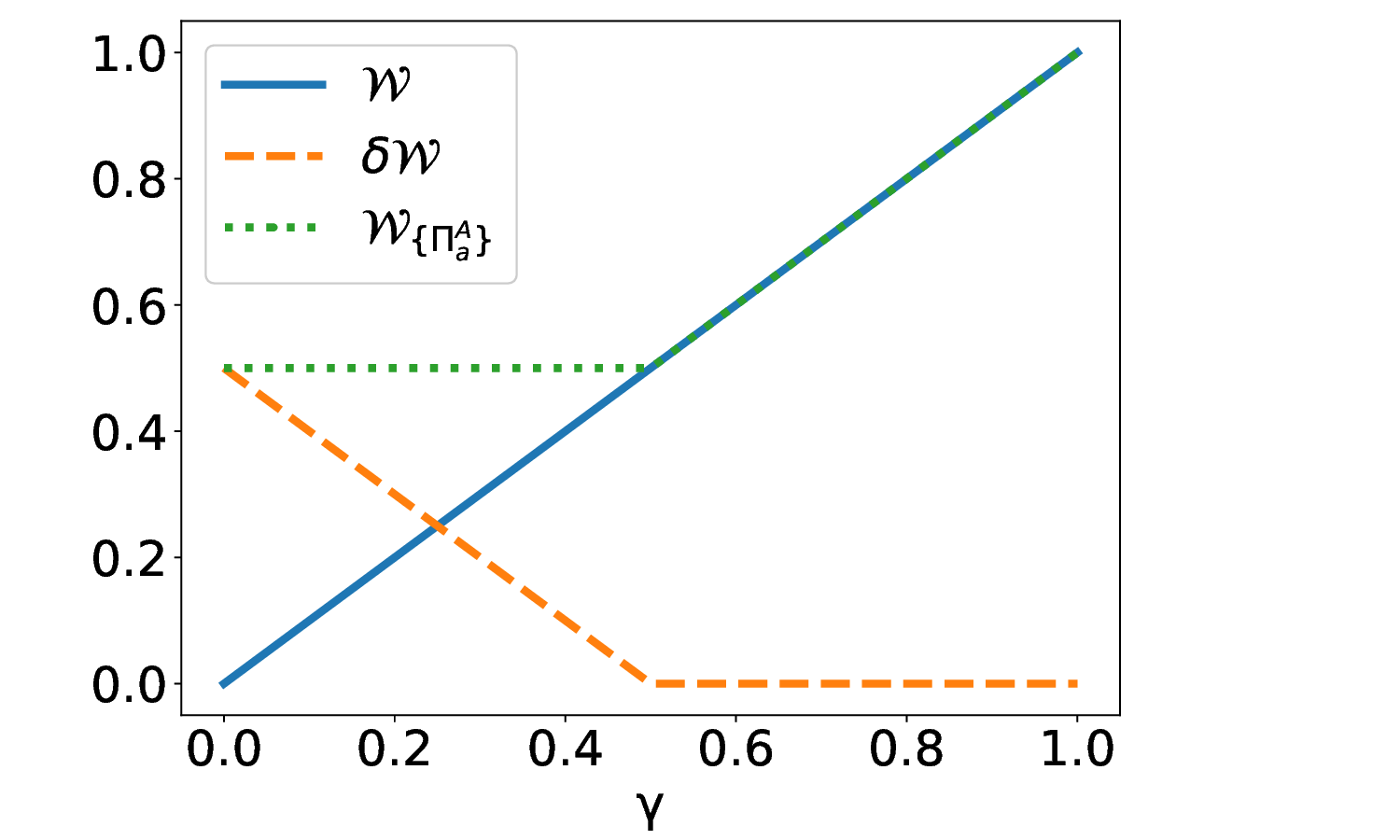}
	\centering
	\caption{Work extraction under amplitude damping as a function of the damping strength \( \gamma \). The solid blue line represents the  ergotropy $\mathcal{W}$,  dotted green line shows the daemonic ergotropy $ \mathcal{W}_{\{\Pi_A\}} $, dashed orange line represents the daemonic gain  $\delta \mathcal{W}$.
 }
	\label{Fig1}
\end{figure}

Fig. (\ref{Fig1}) shows the interplay between ergotropy, daemonic ergotropy, and daemonic gain in a qubit system subjected to an amplitude damping channel, as a function of the damping strength $ \gamma$.

The blue solid line shows the ergotropy $\mathcal{W}$ in terms of damping strength $\gamma$. As the damping strength \( \gamma \) increases, the population in the excited state increases, which results in a larger energy imbalance. This imbalance leads to more extractable work, causing $\mathcal{W}$ to increase linearly with $\gamma$.

 Initially, the system and ancilla are in a classically correlated state in Eq. \ref{cc state} , meaning there is no quantum correlation between them. When local  non-unital noise, such as the amplitude damping channel, is applied to the system, it induces a transformation that generates quantum correlations between the system and the ancilla. As a result of these induced quantum correlations, the system's ergotropy is enhanced. The ergotropy increases because the presence of quantum correlation enables more efficient work extraction. The quantum correlations generated through local noise thus play a crucial role in boosting the extractable work, demonstrating the fundamental relationship between noise, correlations, and work extraction in open quantum systems.
 
It is observed that  for \( \gamma \leq 0.5 \), the daemonic ergotropy exceeds the system ergotropy, i.e., \( \mathcal{W}_{\{\Pi_A\}} > \mathcal{W} \), resulting in a positive daemonic gain \( \delta \mathcal{W} > 0 \). This indicates that the system, in isolation, is energetically passive (or nearly passive), but quantum correlations with the ancilla-generated through the amplitude damping-allow a measurement-assisted protocol to extract more work than is possible through unitary operations alone. In this regime, the measurement effectively activates otherwise inaccessible energy, similar to a quantum Maxwell's demon exploiting quantum information to gain a thermodynamic advantage.

It can also be seen that for $ \gamma > 0.5 $, the system becomes more non-passive as the population in the excited state increases. Consequently, the ergotropy increases and reaches the level of the daemonic ergotropy, i.e., $ \mathcal{W}_{\{\Pi_A\}} = \mathcal{W} $. In this case, the measurement-based protocol provides no additional advantage, and the daemonic gain vanishes, $ \delta \mathcal{W} = 0 $. Physically, this signifies that the system has become sufficiently ordered by the noise itself, rendering the ancilla measurement redundant.
\section{Correlated channel dynamics and their influence on work extraction}\label{sec4}
When studying how quantum systems interact with their surroundings, it is important to consider the nature of the noise they experience. Broadly speaking, quantum channels fall into two main categories: those without memory and those with memory.

In a memoryless channel, each interaction between the system and the environment is treated as completely independent from the last. This is a reasonable assumption when the environment forgets quickly, meaning its internal changes happen much faster than the time between two uses of the channel. In such cases, the environment does not retain any influence from one qubit to the next, so every use of the channel applies the same kind of noise, unaffected by what came before.

These types of channels are simpler to analyze and are commonly used in theoretical studies. They offer a clean picture where each piece of quantum information experiences noise in isolation, making it easier to predict how the system will evolve.

In essence, when a quantum channel is memoryless, the same noise operation is applied to the system each time it passes through the channel. Since no information is retained between successive uses, each application of the channel operates independently. If the channel is used \( N \) times, its overall effect on the system is represented by the tensor product of identical operations, denoted as \( \Lambda^{\otimes N} \). This formalism captures the key characteristic of memoryless noise: it is uncorrelated, consistent, and resets with each use. However, in real-world scenarios, the assumption of independent uses of the channel often fails. Physical environments can retain information about their interactions with the system, particularly when the time between uses of the channel is shorter than the  correlation time of environment. In such situations, the noise acting on one part of the system may depend on previous interactions, leading to correlations between successive uses of the channel. As a result, the overall channel can no longer be described as a simple product of identical operations, i.e., $\Lambda^{\otimes N} \neq \Lambda^N$. This type of behavior is what we refer to as a quantum channel with memory. To explore the behavior of channels with memory, we consider the scenario where a quantum channel is applied \( N \) times to a system. For a given input state \( \rho \), the channel \( \epsilon \) acts as a completely positive, trace-preserving (CPTP) map, which ensures that the quantum evolution remains physically valid. This transformation is typically expressed using a set of Kraus operators \( \{ E_i \} \). The operators \( E_i \), defined as $E_i =  \sqrt{P_{i_1} \dots P_{i_N}} A_i$, represent the Kraus operators of the quantum channel. These operators must satisfy the completeness condition, which is ensured by the normalization of the associated probabilities. Specifically, the sum over all possible sequences of operators must equal one, $\sum P_{i_1 \dots i_N} = 1$. The term \( P_{i_1 \dots i_N} \) represents the probability associated with a specific random sequence of operations being applied to the \( N \) qubits as they pass through the channel. In the case of a memoryless quantum channel, the operations \( A_i \) applied during each use are statistically independent. As a result, the joint probability of a sequence of operations across \( N \) uses factorizes into a simple product $
P_{i_1 \dots i_N} = P_{i_1} P_{i_2} \dots P_{i_N}$. To simplify the analysis, we focus on the case of two consecutive uses of a quantum channel with partial memory. In this scenario, the corresponding Kraus operators can be expressed as

\begin{equation}
E_{i,j} = \sqrt{P_i \left[ (1-\mu) P_j + \mu \delta_{i,j} \right]} K_i \otimes K_j,
\end{equation}
where \( \mu \in [0,1] \) is the memory parameter, and \( \delta_{i,j} \) is the Kronecker delta, ensuring that the same operation is applied to both qubits when \( i = j \). This form of the operator allows the model to interpolate between uncorrelated noise (when \( \mu = 0 \)) and fully correlated noise (when \( \mu = 1 \)). Using the Kraus operator formalism, the evolution of an arbitrary initial state \( \rho \) under a quantum channel with partial memory can be described as:

\begin{equation}
\epsilon(\rho) = (1-\mu) \sum_{i,j} E_{i,j} \rho E_{i,j}^\dagger + \mu \sum_k E_{k,k} \rho E_{k,k}^\dagger,
\end{equation}
Above expression represents a convex combination of two processes: with probability \( 1 - \mu \), the noise operations \( E_{i,j} \) are applied independently to each qubit, while with probability \( \mu \), a correlated operation \( E_{k,k} \) is applied simultaneously to both qubits. The parameter \( \mu \in [0,1] \) governs the degree of memory in the channel, with \( \mu = 0 \) corresponding to no memory (independent noise) and \( \mu = 1 \) indicating full memory (correlated noise). The amplitude damping channel models energy loss, such as spontaneous emission in atomic systems, where a quantum system transitions from an excited state to a lower energy state. This channel is non-unital, as it does not preserve the maximally mixed state. The Kraus operator for the amplitude damping channel is given in Eq.(\ref{kramp}).

Consider the case where two qubits are transmitted through a quantum amplitude damping channel under uncorrelated noise conditions, meaning the environmental correlation time is shorter than the time between successive uses of the channel. In this scenario, each qubit interacts independently with its local environment. The overall noise process can be described using a set of Kraus operators, which are constructed as tensor products of single-qubit operations
\begin{equation}
E_{i,j} = K_i \otimes K_j, \quad \text{for} \quad i,j \in \{0,1\}.
\end{equation}
Above formulation captures the memoryless nature of the channel, where each qubit undergoes its own amplitude damping process independently of the other. As established in earlier works \cite{31, 32}, the Kraus operators \( E_{k,k} \) for an amplitude damping channel with finite memory, acting over two consecutive uses, are given by
\begin{equation}
E_{00}=\left(
\begin{array}{cccc}
 1 & 0 & 0 & 0 \\
 0 & 1 & 0 & 0 \\
 0 & 0 & 1 & 0 \\
 0 & 0 & 0 & \sqrt{1-\gamma } \\
\end{array}
\right), \quad E_{11}=\left(
\begin{array}{cccc}
 0 & 0 & 0 & \sqrt{\gamma } \\
 0 & 0 & 0 & 0 \\
 0 & 0 & 0 & 0 \\
 0 & 0 & 0 & 0 \\
\end{array}
\right).
\end{equation}
Here, we consider the initially classically correlated state, as described in Eq. (\ref{cc state}). After applying the amplitude-damping channel with correlated noise over two consecutive uses, the transmitted density matrix can be obtained as
\begin{eqnarray}
\rho_{SA}'&=&\frac{1}{2} (\gamma  ( \gamma +\mu(1- \gamma) )+1) \vert e \rangle_S \langle e \vert \otimes \vert e \rangle_A \langle e \vert  \nonumber \\
&+& \frac{1}{2} \gamma (\gamma -1)  (\mu -1)\vert e \rangle_S \langle e \vert \otimes \vert g \rangle_A \langle g \vert \nonumber \\
&+& \frac{1}{2} \gamma(\gamma -1)   (\mu -1)\vert g \rangle_S \langle g \vert \otimes \vert e \rangle_A \langle e \vert \nonumber \\
&-& \frac{1}{2} (\gamma -1) (\gamma  (\mu -1)+1)\vert g \rangle_S \langle g \vert \otimes \vert g \rangle_A \langle g \vert.
\end{eqnarray}
After taking the partial trace over the ancilla, the reduced density matrix of the quantum system can be obtained as shown in Eq.(\ref{reds}). So the ergotropy of the system is $\mathcal{W}=\gamma \omega$. The situation changes significantly when we consider daemonic ergotropy, which captures the extractable work under measurement-based feedback control. Following the protocol discussed in the previous section, we assume a projective measurement that is performed on the quantum system locally. Specifically, we use a set of orthogonal projectors, as defined in Eq. (\ref{pms}), to perform a measurement on one qubit, typically interpreted as the quantum system. The conditional state of the system following the measurement outcome $0$ is given by
\begin{equation}
\rho_{S|0} =  \alpha_0 \vert e \rangle_S \langle e \vert + \beta_0  \vert g \rangle_S \langle g \vert,
\end{equation}
where 
\begin{eqnarray}
\alpha_0 &=& \frac{\left( \gamma+\left[ 1-\gamma(2 \gamma(\mu-1)-2 \mu+1)\right]  \cos \theta+1\right)}{2(\gamma \cos  \theta+1)} \nonumber \\
\beta_0 &=& \frac{\left( (\gamma-1)[(2 \gamma(\mu-1)+1) \cos \theta-1]\right)}{2(\gamma \cos \theta+1)} .
\end{eqnarray}
 Similarly, the conditional state of the system after the measurement outcome \(1\) is given by
\begin{equation}
\rho_{S|1} =  \alpha_1 \vert e \rangle_S \langle e \vert 
+ \beta_1 \vert g \rangle_S \langle g \vert,
\end{equation}
where 
\begin{eqnarray}
\alpha_1 &=&  \frac{\left( \left(2 \gamma ^2 (\mu -1)- 2 \gamma  \mu +\gamma -1\right) \cos  \theta +\gamma +1\right)}{2 (1-\gamma  \cos  \theta )}, \nonumber \\
\beta_1&=& \frac{\left( (\gamma -1) ((2 \gamma  (\mu -1)+1) \cos \theta +1)\right)} {2 (1-\gamma  \cos  \theta )}.  
\end{eqnarray}
 After obtaining the conditional states \( \rho_S|0 \) and \( \rho_S|1 \), the next step is to evaluate the ergotropy associated with each state. For a given conditional state \( \rho_S|a \), the ergotropy, denoted as \( \mathcal{W}(\rho_{S|a}) \), can be written as follow
\begin{eqnarray}
\mathcal{W}(\rho_{S|0})&=&\left( 2 \alpha_0 -1\right)  \cdot \Theta \left( \alpha_0 -\frac{1}{2} \right), \\
\mathcal{W}(\rho_{S|1})&=&\left( 2 \alpha_1 -1\right)  \cdot \Theta \left( \alpha_1 -\frac{1}{2} \right). \nonumber
\end{eqnarray}
Ultimately, the  daemonic ergotropy is obtained by taking the weighted average of the ergotropies corresponding to each measurement outcome
\begin{equation}
\mathcal{W}_{\{\Pi_A\}} = p_0 \mathcal{W}(\rho_{S|0}) + p_1 \mathcal{W}(\rho_{S|1}),
\end{equation}
where $p_0=\frac{1}{2} (1+\gamma  \cos ( \theta ))$ and  $p_1=\frac{1}{2} (1-\gamma  \cos ( \theta ))$.
\begin{figure}[H]
	\centering
	\includegraphics[width = \linewidth]{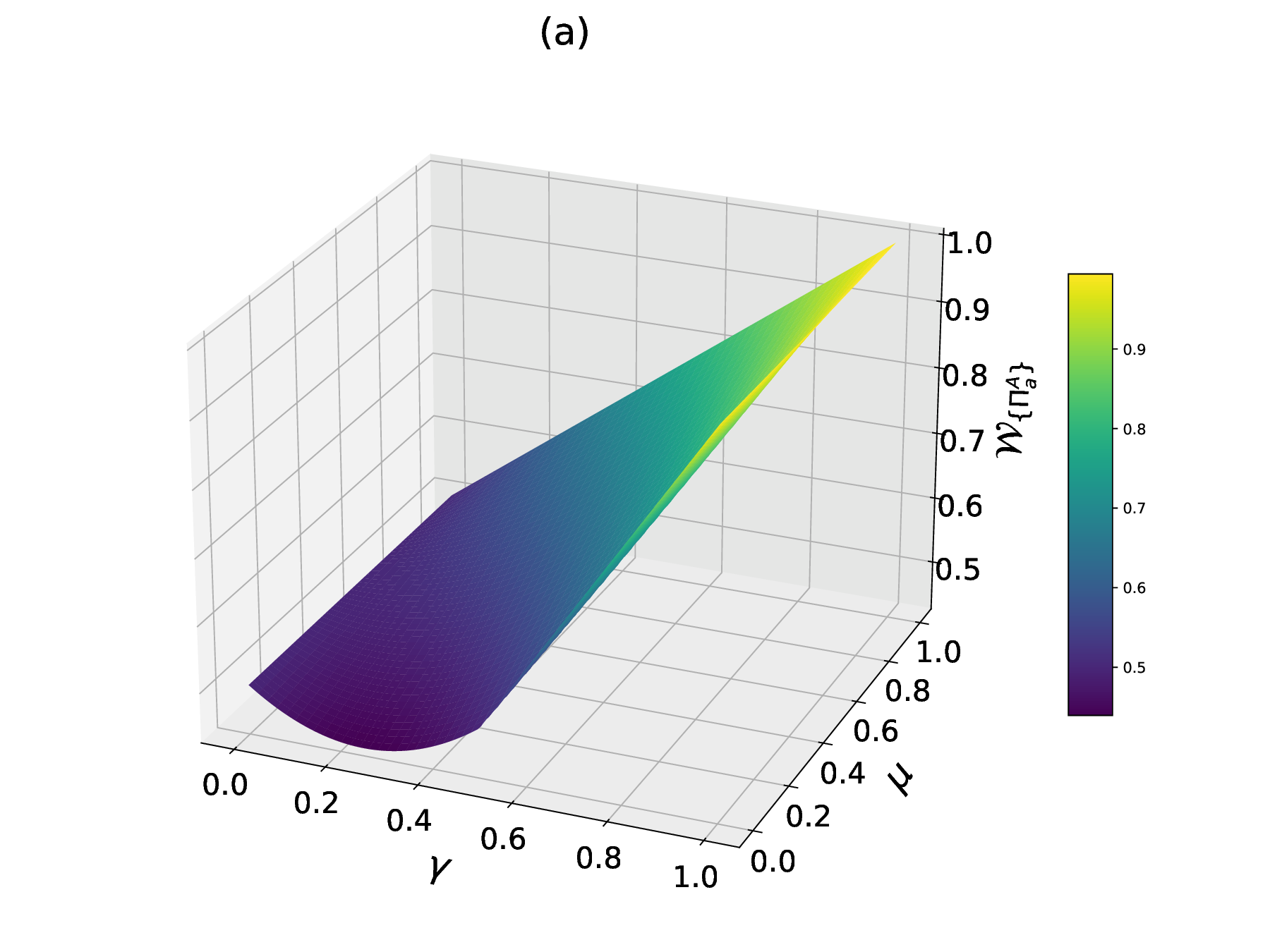}
	\includegraphics[width = \linewidth]{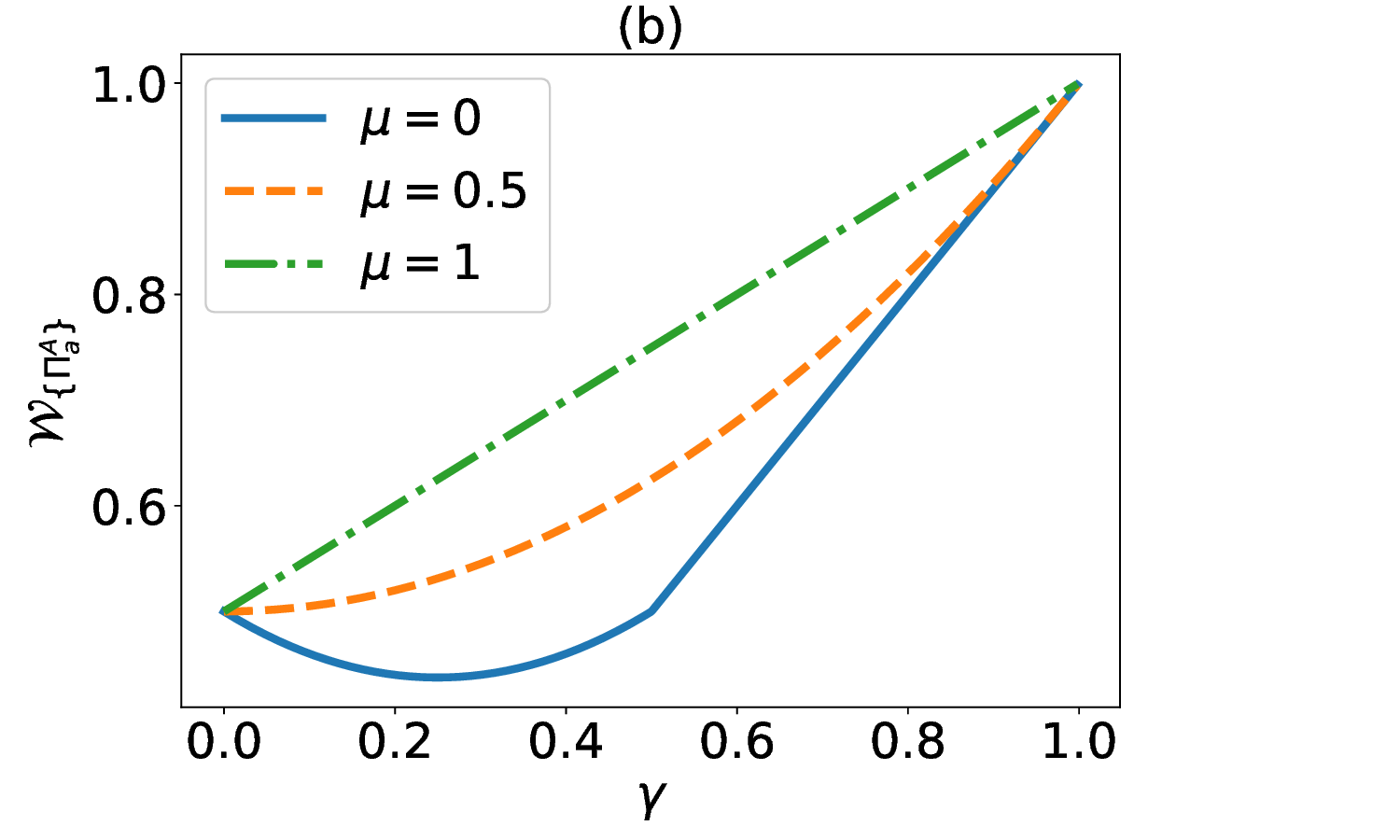}
	\centering
	\caption{\textbf{(a)} The plot of the daemonic ergotropy \( \mathcal{W}_{\{\Pi_A\}} \) as a function of the amplitude damping strength \( \gamma \) and the channel memory coefficient \( \mu \). \textbf{(b)} The plot of \( \mathcal{W}_{\{\Pi_A\}} \) versus \( \gamma \) for different values of \( \mu  \).}
	\label{Fig2}
\end{figure}
Fig. (\ref{Fig2}) shows the behavior of daemonic ergotropy \( \mathcal{W}_{\{\Pi_A\}} \) in a quantum system subjected to an amplitude damping channel with memory effects, characterized by the memory coefficient \( \mu \).  It can be seen from Fig. (\ref{Fig2})(a) that for low \( \mu \) memoryless channels, daemonic ergotropy initially decreases with increasing \( \gamma \), reaching a minimum before rising again. This non-monotonic behavior indicates that for intermediate damping, measurement feedback on the ancilla is less effective due to weaker correlations between the system and the ancilla. As \( \mu \) increases, the surface smooths out and tilts upward, demonstrating that memory in the channel enhances the ability to extract work via daemonic protocols. This improvement occurs because memory effects preserve system-ancilla correlations across successive uses of the channel, thereby enhancing  useful correlations. At high \( \mu \) and high \( \gamma \), the extractable work approaches its  maximum, suggesting that noise with memory can be constructively harnessed for thermodynamic gain.
Fig.(\ref{Fig2})(b) shows slices of the 3D surface at constant \( \mu \), providing a clearer view of how memory affects the \( \gamma \)-dependence of daemonic ergotropy.
It can be seen that the daemonic ergotropy \( \mathcal{W}_{\{\Pi_A\} }\) increases with both the damping strength \( \gamma \) and the memory coefficient \( \mu \). For a memoryless channel (\( \mu = 0 \)), work extraction is initially suppressed for small \( \gamma \). However, as \( \mu \) increases, the ergotropy becomes a monotonic function of \( \gamma \), reaching its maximum at \( \gamma = 1 \). This demonstrates that channel memory enhances the effectiveness of measurement-based feedback, enabling greater work extraction by preserving or amplifying system-ancilla correlations. 
\begin{figure}[H]
	\centering
	\includegraphics[width = \linewidth]{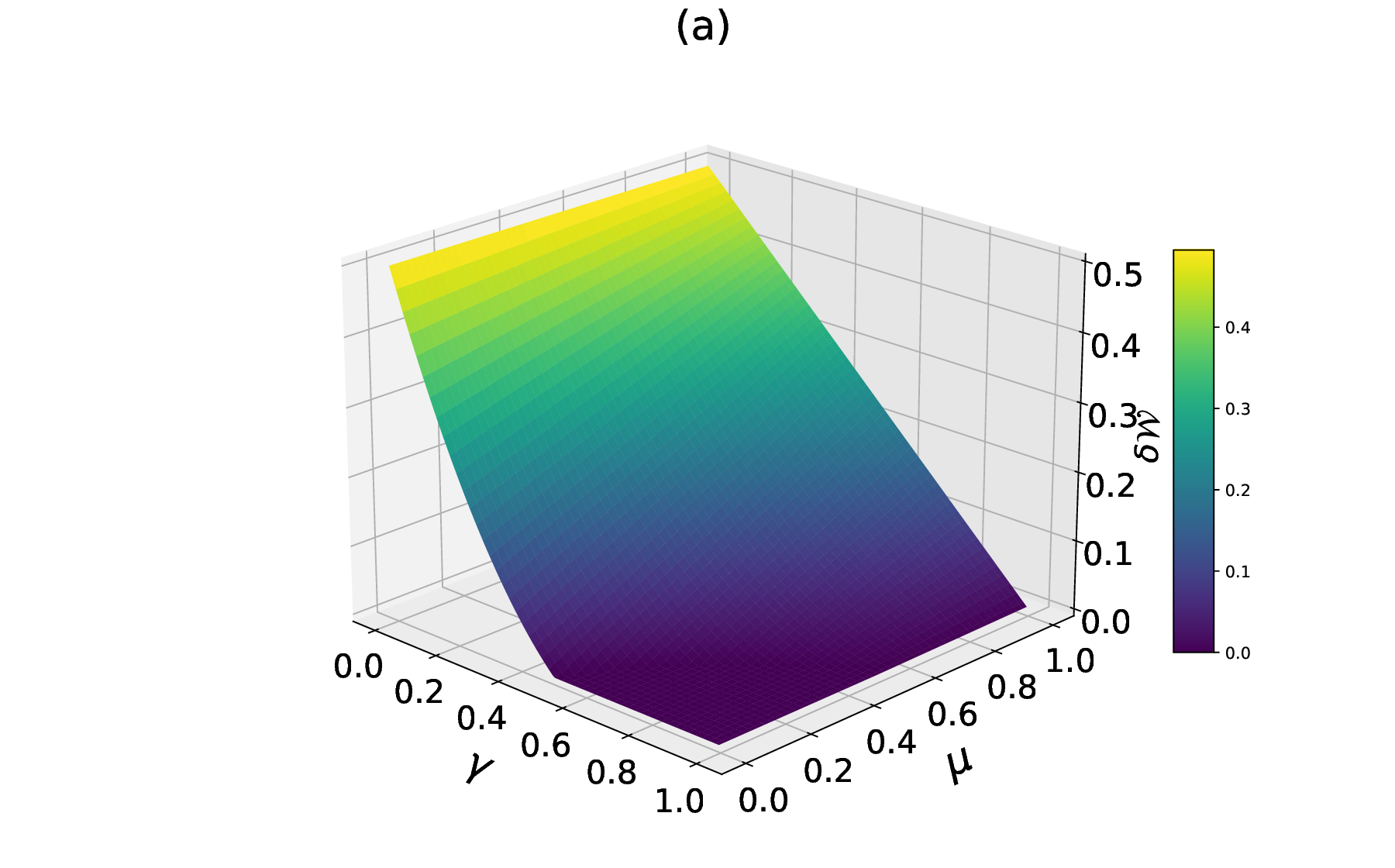}
	\includegraphics[width = \linewidth]{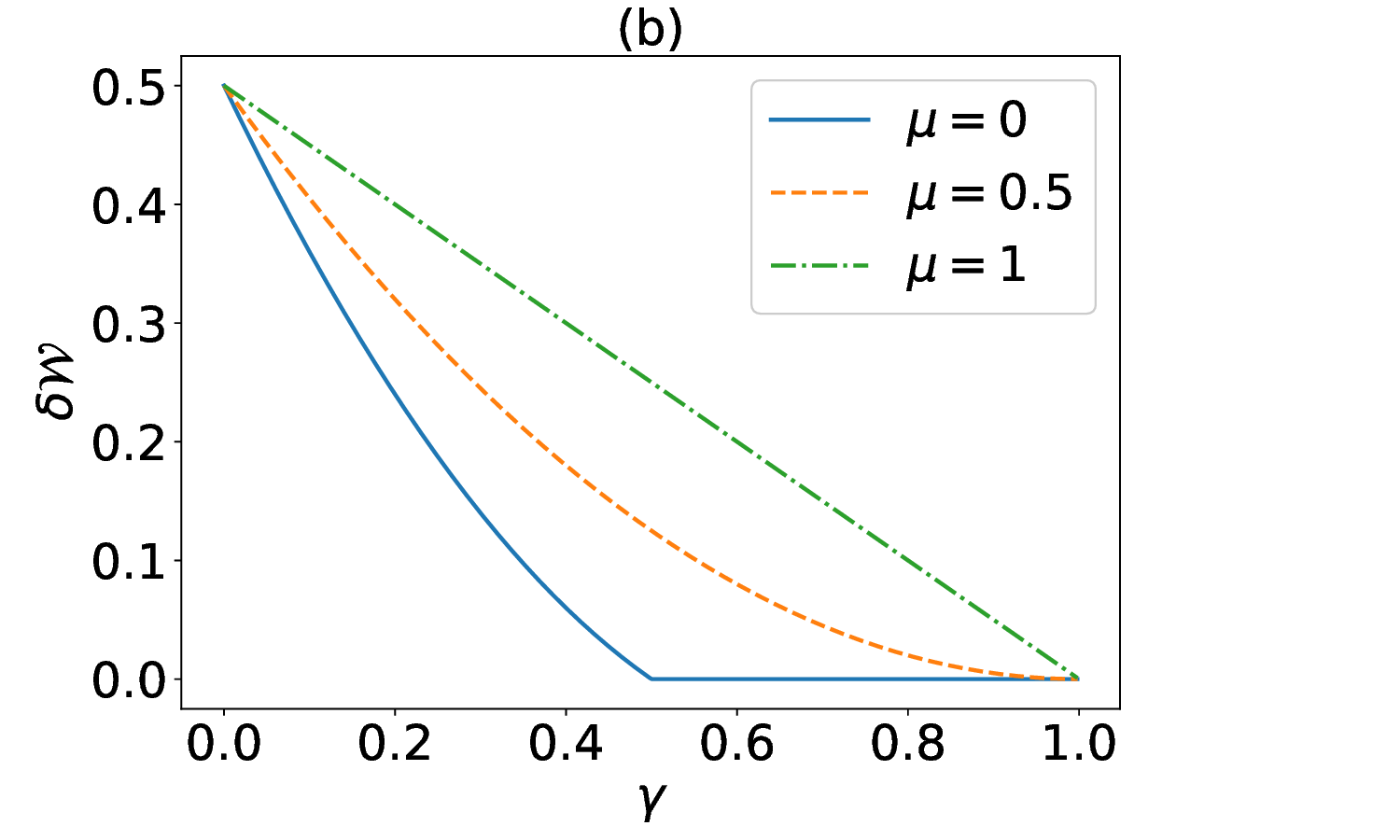}
	\centering
	\caption{\textbf{(a)} The plot of the daemonic gain \( \delta \mathcal{W} =\mathcal{W}_{\{\Pi_A\}} - \mathcal{W} \) as a function of the damping strength \( \gamma \) and the channel memory coefficient \( \mu \). 
\textbf{(b)} The plots of \( \delta W \) versus \( \gamma \) for fixed values of \( \mu = 0 \), \( \mu = 0.5 \), and \( \mu = 1 \). 
}
	\label{Fig3}
\end{figure}
Fig. \ref{Fig3} illustrates the behavior of the daemonic gain \( \delta W =  \mathcal{W}_{\{\Pi_A\}}  - \mathcal{W} \), under correlated amplitude damping channel. The results are shown as a function of the amplitude damping strength \( \gamma \) and the channel memory coefficient \( \mu \), where \( \mu = 0 \) corresponds to a uncorrelated channel, and \( \mu = 1 \) represents a fully correlated channel.

Fig. \ref{Fig3}(a) shows the plot of \( \delta W \) as function of $\gamma$ and $\mu$. It is evident that the daemonic gain decreases monotonically with increasing \( \gamma \), reflecting the fact that stronger decoherence reduces the quantum correlations necessary to boost work extraction. However, as \( \mu \) increases, the daemonic gain is significantly enhanced and sustained over a broader range of \( \gamma \). This suggests that channel memory helps preserve, or even regenerate, useful quantum correlations, allowing the feedback-assisted protocol to remain effective despite stronger noise.

Fig. \ref{Fig3}(b) presents cross-sectional views of \( \delta W \) as a function of \( \gamma \) for different values of \( \mu \). For a memoryless channel (\( \mu = 0 \)), the daemonic gain rapidly drops to zero around \( \gamma = 0.5 \). In contrast, for \( \mu = 0.5 \) and \( \mu = 1 \), the daemonic gain decays more gradually. The \( \mu = 1 \) case shows a nearly linear decrease, remaining nonzero until \( \gamma = 1 \). This highlights the constructive role of memory: it counteracts the destructive effects of decoherence and extends the thermodynamic utility of ancilla-assisted protocols.

These observations align with the findings of Streltsov et al. \cite{Streltsov2011}, which showed that non-unital local channels, such as amplitude damping, can generate quantum correlation from initially classically correlated states. Additionally, memory effects help preserve such quantum correlation over time. Within the framework of daemonic ergotropy, memory of the channel thus acts as a thermodynamic resource that enhances the operational usefulness of quantum measurements. Specifically, the memory-assisted retention of correlations allows the quantum demon to remain effective even in regimes where noise would otherwise suppress extractable work.

In summary, the presence of memory in the amplitude damping channel not only boosts the daemonic gain but also extends the regime in which quantum feedback mechanisms outperform purely unitary extraction. This demonstrates that memory of the non-unital channel can be harnessed as a powerful enabler for energy extraction in quantum thermodynamic systems.
\begin{figure}[H]
	\centering
	\includegraphics[width = \linewidth]{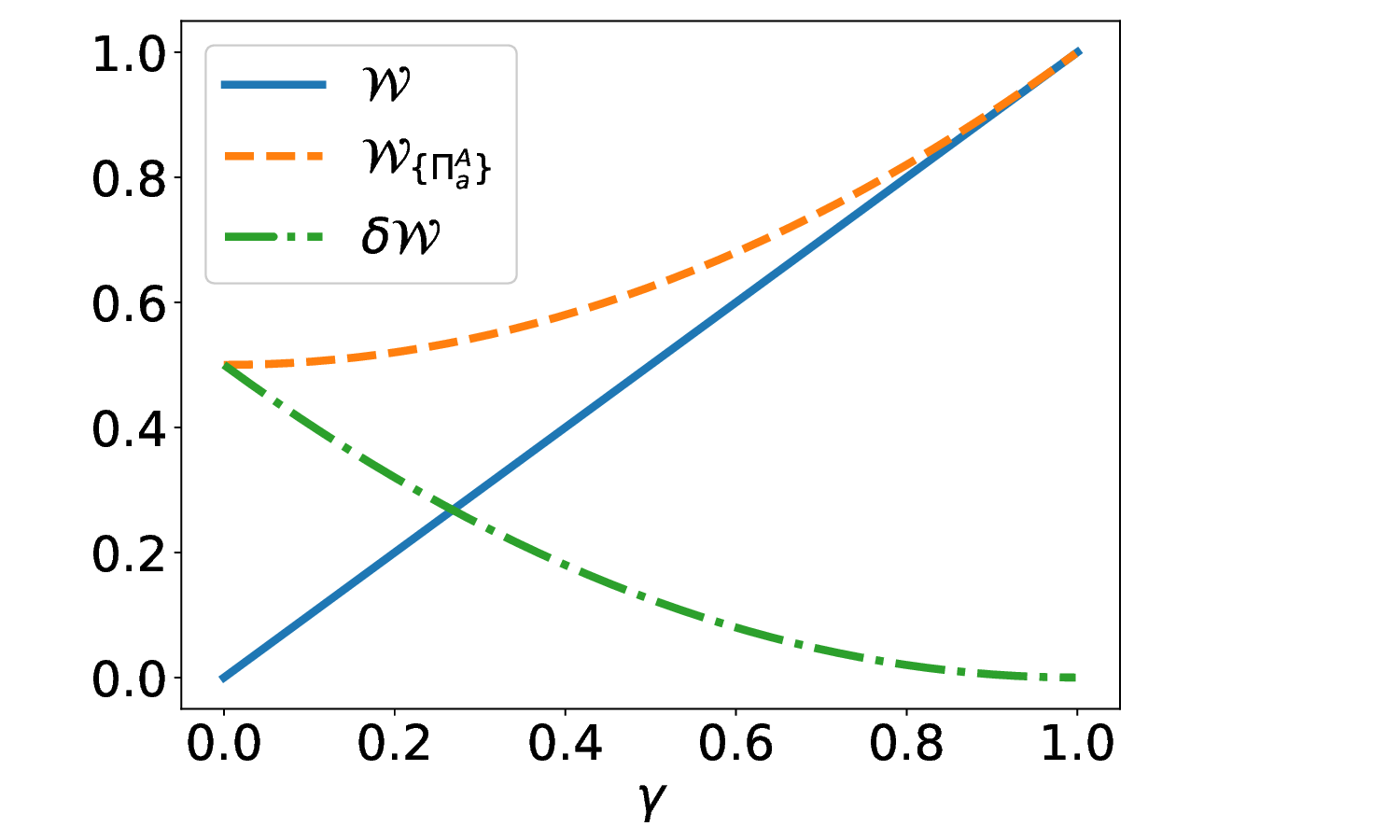}
	\centering
	\caption{The ergotropy \( \mathcal{W} \), daemonic ergotropy \( \mathcal{W}_{\{\Pi_A\}} \), and daemonic gain \( \delta \mathcal{W} \) are shown as functions of the amplitude damping strength \( \gamma \) for a fixed memory coefficient \( \mu = 0.5 \).
}
	\label{Fig4}
\end{figure}
Fig. \ref{Fig4} displays the behavior of three key quantities as functions of the amplitude damping strength \( \gamma \), with the channel memory parameter fixed at \( \mu = 0.5 \). As can be seen ergotropy $\mathcal{W}$ increases linearly with $\gamma$. This aligns with the standard thermodynamic picture where a more asymmetric state allows more work extraction. 

It can also be seen that the daemonic ergotropy \( \mathcal{W}_{\{\Pi_A\}} \) grows more rapidly and nonlinearly, consistently staying above the ergotropy curve. This behavior indicates that measurement-based feedback on the ancilla allows access to additional structured energy in the system, which arises due to the presence of quantum correlations between system and ancilla.

In Fig.\ref{Fig4}, \( \delta W \) remains strictly non-zero across the entire range of damping strengths \( \gamma \) of correlated channel, signaling that the measurement-based protocol continues to outperform purely unitary work extraction, even as decoherence increases. Physically, this non-zero gain suggests that the system retains useful quantum correlations with the ancilla, which can be exploited by the demon (i.e., the measurement-feedback protocol) to enhance thermodynamic performance. These correlations are not accessible through local operations alone, but become operational when the measurement is performed, and the resulting conditional states are used for work extraction. The presence of memory (\( \mu = 0.5 \)) plays a crucial role here. it helps preserve these correlations during the evolution, preventing their rapid decay under noise. As a result, the demon remains effective across a broader range of damping, enabling a quantum advantage in work extraction. In essence, the non-zero daemonic gain reflects the operational power of quantum discord and its preservation through memory effects, demonstrating that, even in noisy environments, structured measurements can unlock hidden thermodynamic value from correlations that would otherwise be inaccessible.
\section{Conclusion}\label{sec5}
In this work, we have explored the role of local noise and channel memory in enabling work extraction from classically correlated quantum systems. Using the framework of daemonic ergotropy, we investigated how a local amplitude damping channel, typically associated with energy loss, can actually act as a source of useful quantum correlations when applied to one part of an initially classically correlated bipartite state. Our analysis demonstrated that such non-unital dynamics can induce nonclassical correlations  even in the absence of entanglement, thereby activating the potential for enhanced thermodynamic performance.

By considering projective measurements on an ancilla qubit and applying feedback operations to the system, we quantified the enhancement in extractable work through daemonic ergotropy. Our results confirmed that, although the system starts in a passive configuration with zero ergotropy, the action of the local amplitude damping channel on the reduced system renders it non-passive, opening the door to non-zero work extraction. Moreover, we showed that ancilla-assisted protocols yield strictly greater work than unassisted unitaries, as quantified by the daemonic gain. This highlights the operational value of measurement-based feedback in extracting work from noisy, yet correlated, quantum systems.

A central focus of our study was the influence of memory effects, modeled via a correlated amplitude damping channel characterized by a memory coefficient \( \mu \). We found that as \( \mu \) increases, the channel better preserves quantum correlations across uses, thereby sustaining the effectiveness of the daemonic protocol over a broader range of damping strengths. In particular, for intermediate values of \( \gamma \), where the thermodynamic advantage typically diminishes in memoryless scenarios, the presence of memory revives and even enhances the daemonic gain. This indicates that non-Markovianity, often regarded as an unavoidable complexity, can actually be harnessed as a functional resource in quantum thermodynamic tasks.

Overall, our findings establish that the interplay between local noise, memory effects, and quantum measurements can unlock hidden energetic potential in classically correlated systems. This contributes to the growing understanding that noise, when structured and exploited appropriately, can be transformed from an adversary into an ally in quantum technologies. The insights gained from this study not only deepen the thermodynamic interpretation of quantum correlations but also offer practical guidelines for designing energy-efficient quantum devices that actively leverage their environments.

\appendix*

\end{document}